\documentclass[journal]{IEEEtran}

\usepackage[shortlabels]{enumitem}

\usepackage{amsmath,amsfonts,amssymb}
\usepackage{epsfig}
\usepackage[hyphens]{url}
\usepackage{graphicx,color,footmisc}
\usepackage{subcaption}
\usepackage{algorithm}
\usepackage{cite}

\title{Connecting the Unconnected: Towards Frugal 5G Network Architecture and Standardization}
\author{\IEEEauthorblockN{Meghna Khaturia\IEEEauthorrefmark{1}, Pranav Jha\IEEEauthorrefmark{1} and Abhay Karandikar\IEEEauthorrefmark{1}\IEEEauthorrefmark{2} \\}
	\IEEEauthorblockA{\IEEEauthorrefmark{1}Department of Electrical Engineering, Indian Institute of Technology Bombay, Mumbai, India\\ Email: \{meghnak, pranavjha, karandi\}@ee.iitb.ac.in \\}
	\IEEEauthorblockA{\IEEEauthorrefmark{2}Director and Professor, Indian Institute of Technology, Kanpur, India \\ Email: karandi@iitk.ac.in}}
\date{}

\begin{document}
	\maketitle
	
	\begin{abstract}
    This article adopts a holistic approach to address the problem of poor broadband connectivity in rural areas by suggesting a novel wireless network architecture, also called the ``Frugal 5G Network". To arrive at the Frugal 5G Network architecture, we take into consideration the rural connectivity needs and the characteristics specific to rural areas. As part of the proposed Frugal 5G Network, we define a heterogeneous Access Network wherein macro cells provide a carpet coverage while Wireless Local Area Networks (WLANs) provide additional capacity to serve the village clusters. WLAN is backhauled via a wireless network also called the wireless middle mile network. We define a Software Defined Networking (SDN) and Network Function Virtualization (NFV) based architecture to make the network flexible and scalable. The concepts of Fog computing have also been employed in the network architecture to bring intelligence to the edge, i.e., to the access network. Through a novel amalgamation of these technologies, we are able to address the connectivity requirements of rural areas.  The proposed network architecture can serve as a potential solution towards IEEE P2061, a standardization project that aims to design an architecture to facilitate rural broadband communication.
	\end{abstract}

\section{Introduction}
    The importance of Internet in our lives today cannot be overemphasized. So much so that access to Internet has been declared a fundamental right of citizens in several countries such as Finland, Spain, and Greece~\cite{rtinternet}. In most of the developed countries, high-speed Internet connectivity is enabled through wired communication infrastructure such as Fiber-to-the-Home (FTTH) and Very high bit-rate Digital Subscriber Line (VDSL). Unfortunately, the situation is not particularly encouraging in developing countries due to non-availability of such a pervasive communication infrastructure. The difference in the fiber-deployed-to-population ratio across developed and developing countries further underscores this disparity; while this ratio is 1.2 in USA, it is barely 0.1 in India~\cite{deloitte}. Owing to this inadequacy of fiber/DSL availability, cellular access technology has emerged as the primary broadband access mechanism in developing countries. However, the penetration of cellular network is limited in rural areas as its deployment becomes unviable due to challenges such as low average revenue per user, sparse population density, and intermittent availability of electricity. This situation leaves majority of the rural people unconnected thereby creating a massive rural-urban digital divide. The next generation cellular system can bridge this divide if we overcome the above mentioned challenges. However, since the Fifth Generation (5G) cellular technology has focus on requirements such as $10$ Gbps data rate, $1$ ms latency, and $500$ km/h mobility, the problems of coverage and affordability are likely to persist, further widening the digital divide~\cite{IMT2020}.
    
    
    
    In the existing literature, researches have suggested many schemes to enable a broadband network suited for rural areas. Chiaraviglio et al. have examined the possibility of large cells and mounting Remote Radio Heads on the unmanned aerial vehicles~\cite{chiaraviglio20165g}. Economic analysis of these techniques show that the monthly subscription fee can be kept sufficiently low. Karlsson et al. have analyzed the impact of beamforming and cell discontinuous transmission on energy consumption of the network~\cite{karlsson2016energy}. They have suggested that beamforming can be a key enabler to significantly reduce power consumption and thereby decrease the  operational cost. Khalil et al. have proposed the usage of TV White Space (TVWS) spectrum with the 5G infrastructure for providing coverage in rural areas~\cite{khalil2017feasibility}. They have formulated an optimization problem to minimize the capital and operational expenses of the proposed system. Our research team has also deployed a large scale TVWS testbed. By conducting various experiments, the feasibility of providing broadband in rural areas using TVWS has been examined~\cite{kumar}.  However, the above-mentioned approaches address the issue of rural broadband connectivity in a piecemeal fashion. In the light of the current scenario, a  comprehensive approach to address this problem is required. We need a bottom-up approach that identifies the challenges related to rural connectivity and proposes a novel network architecture that specifically deals with these challenges and helps connect the unconnected.
    
    Recently, some significant standardization activities have been initiated which aim to develop innovative architectures to facilitate broadband communication and Internet access in rural areas. Internet Research Task Force (IRTF) has started a project called Global Access to the Internet for All (GAIA)  that aims at providing universal and affordable Internet access to everyone, everywhere~\cite{sathiaseelan2014researching}. The GAIA research group has identified various pillars for a network architecture for providing universal access. They have studied the idea of approximate networking, i.e., designing flexible protocols and architecture which can trade off service quality for affordability/accessibility. IEEE has also initiated a working group called Frugal 5G Network~\cite{P2061}. The standard development project P2061\footnote{One of the authors of this paper chairs the IEEE P2061 working group.}, under this working group, aims to design an architecture for a low-mobility, energy-efficient network for providing affordable broadband access in rural areas. Under the P2061 project, it is envisaged that the proposed network would  comprise of a wireless middle mile network, an access network and its associated control and management functions. Our proposition here is closely aligned with the design objectives of IEEE P2061. 
    
    In this paper, we present a novel communication network architecture for rural areas based on our learnings from two large-scale testbeds deployed by our team in Palghar, Maharashtra, India~\cite{kumar,bookchapter}. We consider the rural characteristics and requirements while designing the network and refer to this network as the ``Frugal 5G Network". The proposed architecture presents a framework for the Frugal 5G network which can be realized by utilizing components from the evolving 5G cellular standards and other existing wireless technology standards. The network architecture utilizes the Software Defined Networking (SDN) and Network Function Virtualization (NFV) paradigms, the cornerstones of the emerging 5G cellular technology. It also exploits the Cloud and Fog architectural model consisting of a Cloud-based Core Network and Fog-based Access Network. The AN is heterogeneous in nature, comprising of Macro Base Stations (BS), which could be based on 5G New Radio, and Wireless Local Area Networks (WLANs). The WLANs may be based on IEEE 802.11 Wi-Fi. The backhaul for the WLAN Access Points (APs) is enabled via a wireless multi-hop middle mile network.  The main contribution of this paper is in suggesting a unique and novel combination of these technologies under a unified framework to realize advantages like cost reduction, energy efficiency, rapid service conception, ease of deployment, multi-tenancy, and geography based service introduction thereby addressing the rural broadband communication needs effectively.
    
    
    
    The rest of the paper is organized as follows.  We discuss some of the rural area specific characteristics next. Section III provides an overview of the proposed network architecture. The main components and the working principles of the architecture are described in section IV.  The open issues and the challenges are discussed subsequently. Finally, we present the summary and conclusion of the work.

\section{Characteristics and Challenges of Connectivity in Rural Areas}

    \label{sec:char}
    Considering the poor broadband penetration in rural areas, especially in those of developing countries, it is imperative to first understand the unique traits that characterize these regions and the key challenges that impede the deployment of a broadband network there. In this section, we identify some of these important characteristics and challenges that may influence the architecture and design of such a network.
    
	\begin{enumerate}
	    \begin{figure}
	        \centering
	        \includegraphics[scale = 0.4,trim={0 0 5.75cm 0},clip]{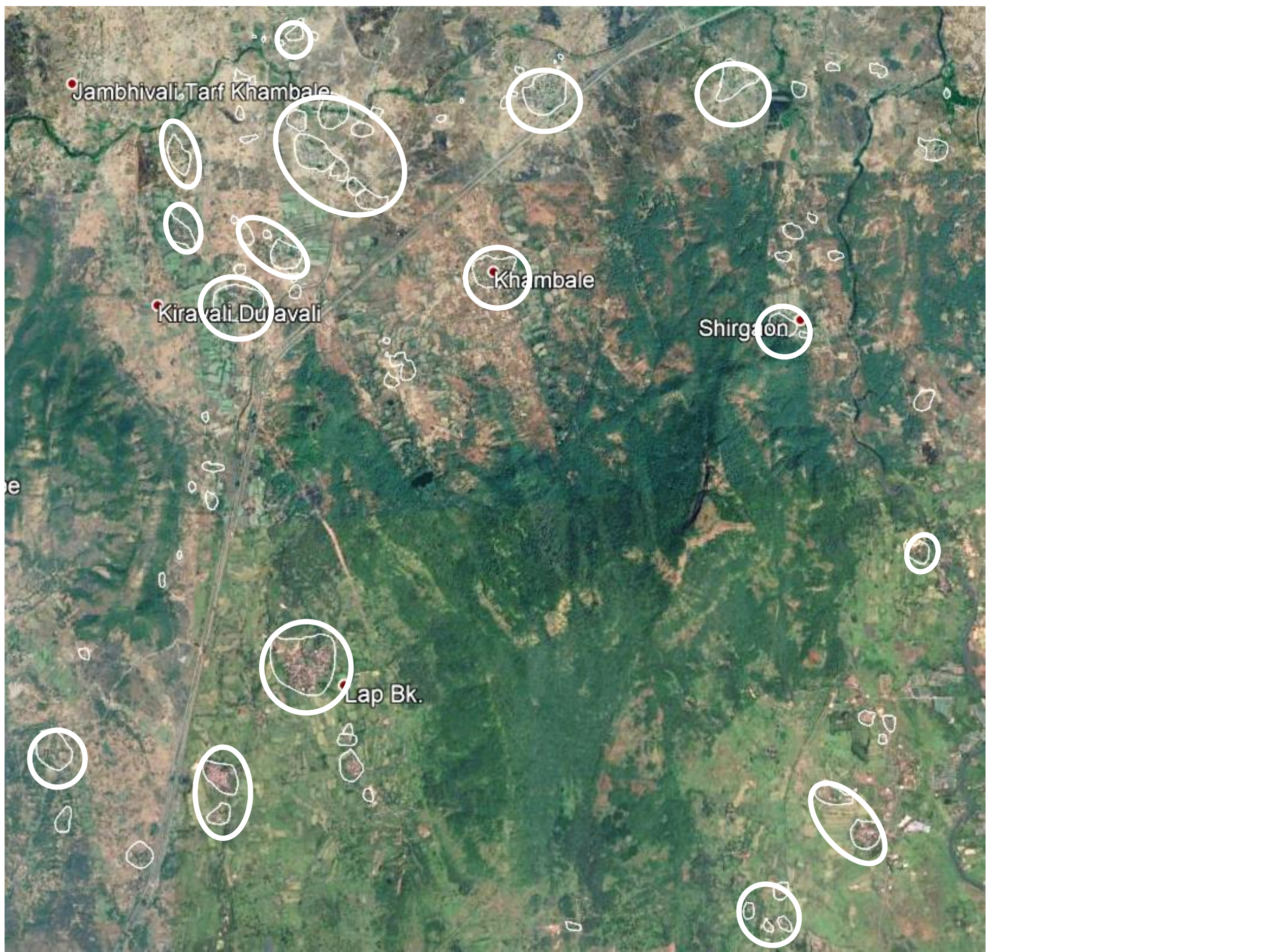}
	        \caption{An example of rural area (approx. 50 km$^2$) depicting the clustered settlement. An ellipse denotes a habitat.}
	        \label{fig:rural}
	        \vspace{-0.5cm}
	    \end{figure}
	    
	         \begin{figure*}
        	\centering
        	\includegraphics[scale=0.45,trim={0 2.5cm 3cm 2.3cm},clip]{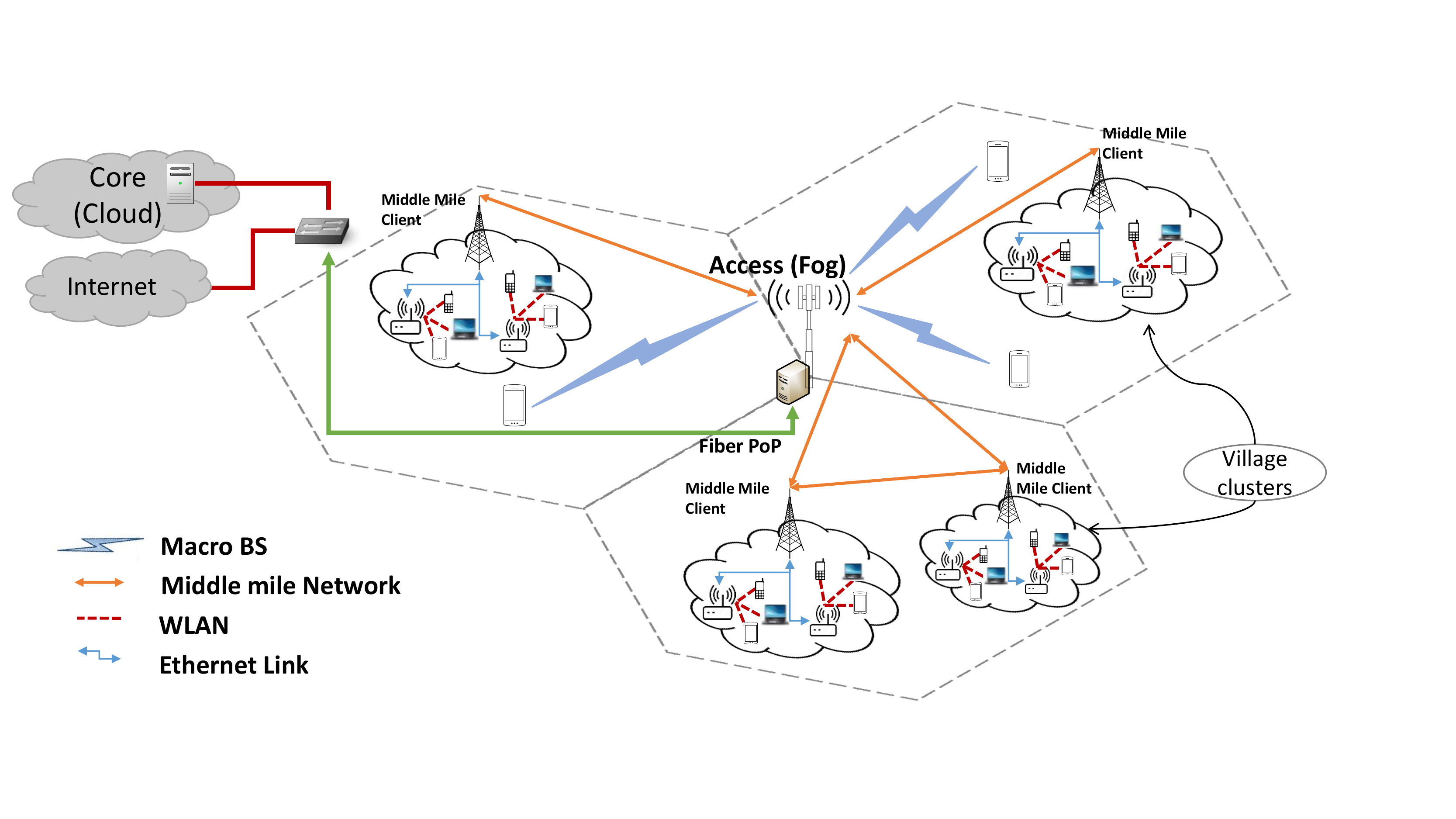}	
        	\caption{A deployment example of Frugal 5G architecture.}
        	\label{fig:dep}
        	\vspace{-0.5cm}
    \end{figure*}
	    \item \textit{Sparse Population Density and Clustered Settlements:} 
	    Unlike urban areas which are uniformly populated, rural areas are populated in clusters which are generally far apart from each other with vast open areas in between as shown in Fig~\ref{fig:rural}. This implies that providing similar type of services over a large geographical area may be economically inefficient. 
        
	    \item \textit{Strong Sense of Community:} One of the important observations from our Palghar testbed is that the people in rural areas have a strong community bond. This suggests that their communication pattern may have a distinct localized nature and a significant fraction of the communication, especially peer-to-peer communication, may be between people living in close vicinity of each other. 
	    
	    \item \textit{Relevance of content:} Today $80$\% of online content is available in $1$ of $10$ languages which only $3$ billion people speak~\cite{2016internet}. In majority of rural areas this $80$\% content has little or no relevance. Additionally, there still exists a gap to facilitate content generation in regional languages for consumption. 
	   
	   \item \textit{Intermittent Availability of Electricity:} In rural areas of developing countries like India, the availability of electricity from the grid is intermittent in nature~\cite{electrification}. It is therefore impossible to guarantee a steady Internet service if the network equipment is solely dependent on electric grid. Hence, it is important to focus on development of technologies capable of running on renewable energy sources in a cost-effective manner.
	   
	   \item \textit{Low Income:} It has been observed that the average income of people living in rural areas is typically much less than that in urban areas. This income disparity is worse in developing countries. For example, the average monthly income of a rural household in India is nearly $\$112$~\cite{secc}, less than $50$\% of an average urban household in the country. 
	    
	    \item \textit{Low Mobility:} People in rural areas, especially in developing countries like India, are either pedestrian or move at low speeds, typically lower than $50$ km/h. In such a scenario, supporting high-speed mobility (up to $500$ km/h) may not have much relevance. Support for high mobility poses various challenges on the design of a cellular system such as developing highly complex channel estimation and synchronization techniques~\cite{mobility}. 
	    
	    \item \textit{Remote/Inaccessible Areas:}  Many parts of the rural areas may be remote and not easily accessible. The maintainability and the availability of the network equipment deployed in such remote regions may be of significant concern. 
	\end{enumerate}
    
	The above-mentioned characteristics highlight various requirements that rural communication network must be based on and are enumerated as follows: i) affordability, ii) limited mobility support, iii) localized communication, iv) capable of running on renewable energy, v) device/equipment security, vi) clustered service provision and vii) facilitation of local content generation and storage.

\section{Frugal 5G: Architectural Overview}
   
   In this section, we propose a method for architecting the standard components of a communication network so as to enable a broadband access network that is calibrated according to rural connectivity requirements. We consider a tiered wireless communication network involving a \textit{Core Network (CN)} and an \textit{Access Network (AN)} as shown in Fig~\ref{fig:dep}. AN facilitates last mile connectivity to end-users, whereas CN provides connectivity with the external data network. In some scenarios, AN may also provide connectivity with the external data network. AN, considered here, is heterogeneous in nature, i.e., it consists of a \textit{Macro} BS and a number of WLANs. We assume that a Point of Presence (PoP) is available in the vicinity of the given rural area. The Macro BS is co-located with the PoP and provides blanket coverage to the end-users in a large geographic area. It also provides the required Quality of Service (QoS) for the critical applications. Owing to the clustered settlement of people in rural areas, WLAN is deployed only in clusters.  IEEE 802.11, the most commonly used WLAN technology, can be chosen for the last mile as it facilitates fixed high-speed broadband access to the end-users in a cost-effective manner.  The major challenge in designing the AN is to backhaul the traffic generated by WLAN APs. This is enabled via wireless Middle Mile Network (MMN) that connects the WLAN APs to the PoP. The middle-mile Access Point (AP) is located near the PoP which wirelessly communicates with the middle mile clients (located in the village clusters) which in turn serve the WLAN APs. The MMN supports optimized mutli-hop mesh network topology to effectively address the network coverage. The architecture proposed here (Fig~\ref{fig:dep}) mirrors the rural population distribution (Fig~\ref{fig:rural}) and helps in efficient delivery of services to the users. In order to control and manage this heterogeneous network in a unified and efficient manner, we exploit three state-of-the-art network architecture and design paradigms---SDN, NFV and Fog Computing \& Communication.

    Following the SDN paradigm, the Frugal 5G network is divided into separate control and data (forwarding) plane elements. The separation of control and data plane functionality is done both in the AN as well as in the CN. While the individual data plane entities of all three components of the AN, i.e., the WLAN, the Macro BS, and the Wireless MMN may be distributed, the control logic of all three components is brought together to facilitate a \textit{unified control and management structure}.  This unified control brings efficiency in the delivery of services. Depending on the service requirements and the network condition, the individual data flows are dynamically setup through appropriate AN data plane elements, for example, either through the MMN and the WLAN or through the Macro BS.    

    All the control and data plane elements in the Frugal 5G network are instantiated as virtual network functions which can run on commercial off-the-shelf platforms. Typically, a cellular network is upgraded every 5 years to keep up with the technological advances. If virtualization is enabled, these upgrades will happen only in software, keeping the physical infrastructure intact, consequently reducing the capital expenditure. This may also aid operators in enabling regional services which would, otherwise, have been difficult to implement on dedicated hardware. 
    
    Owing to the strong community relations and the need to facilitate local content generation and consumption in rural areas, we enable localized communication in our proposition. Towards this, we propose the usage of Fog Computing and Communication. Fog is a smaller Cloud which is closer to the user (in the edge) and brings intelligence to the edge. The deployment of intelligence in the edge enables more efficient communication with reduced resource usage, such as, routing end-to-end data flows locally within the AN wherever applicable, cost-effective delivery of relevant regional content from cache servers in the fog, enabling local communication even when connectivity with CN or external network is unavailable.       
    
    \begin{figure}[!ht]
    		\centering
    		\includegraphics[width = 0.8\linewidth, trim = {0 0 4.8cm 0}, clip]{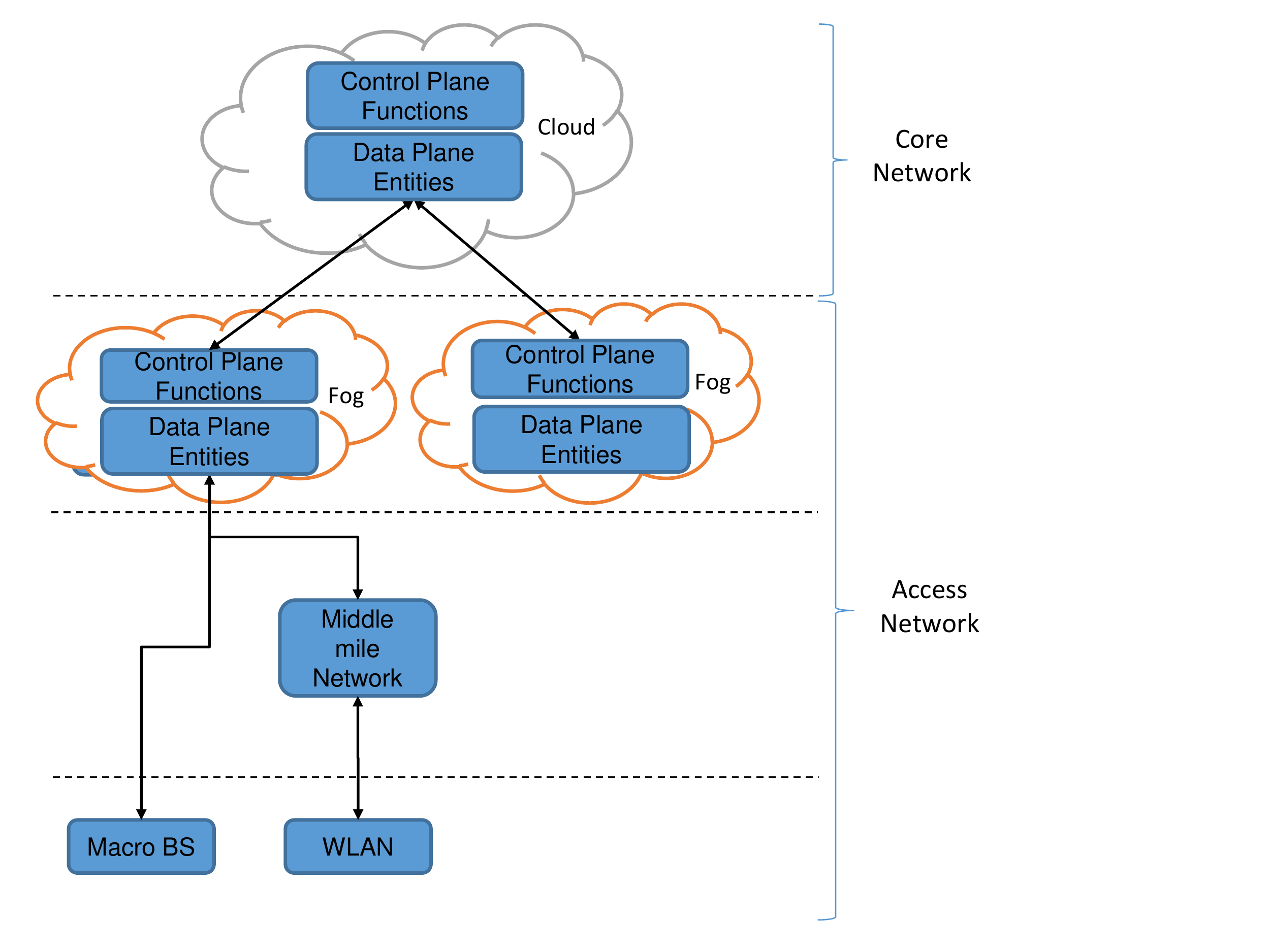}    		\caption{\label{fig:concept} Frugal 5G Architecture Concept}
    	\end{figure}
    
    Fig~\ref{fig:concept} provides an essence of the proposed network architecture and its elements. The proposed architecture possesses a hierarchical structure comprising of a centralized Cloud element and multiple distributed Fog elements located near the users. The Fog and the Cloud, both, use SDN based architecture with well-defined separation between control and data plane entities. The centralized Cloud hosts the CN and performs a set of functions which includes the ones performed by traditional wireless core networks. A Fog comprises of AN elements belonging to different Radio Access Technologies (RATs). A controller in each Fog element controls and manages data plane entities from multiple RATs.
    

\section{Main Components}
    \begin{figure*}[!ht]
		\centering
		\includegraphics[height = 8cm, trim={0cm 0cm 0.5cm 0.5cm},clip]{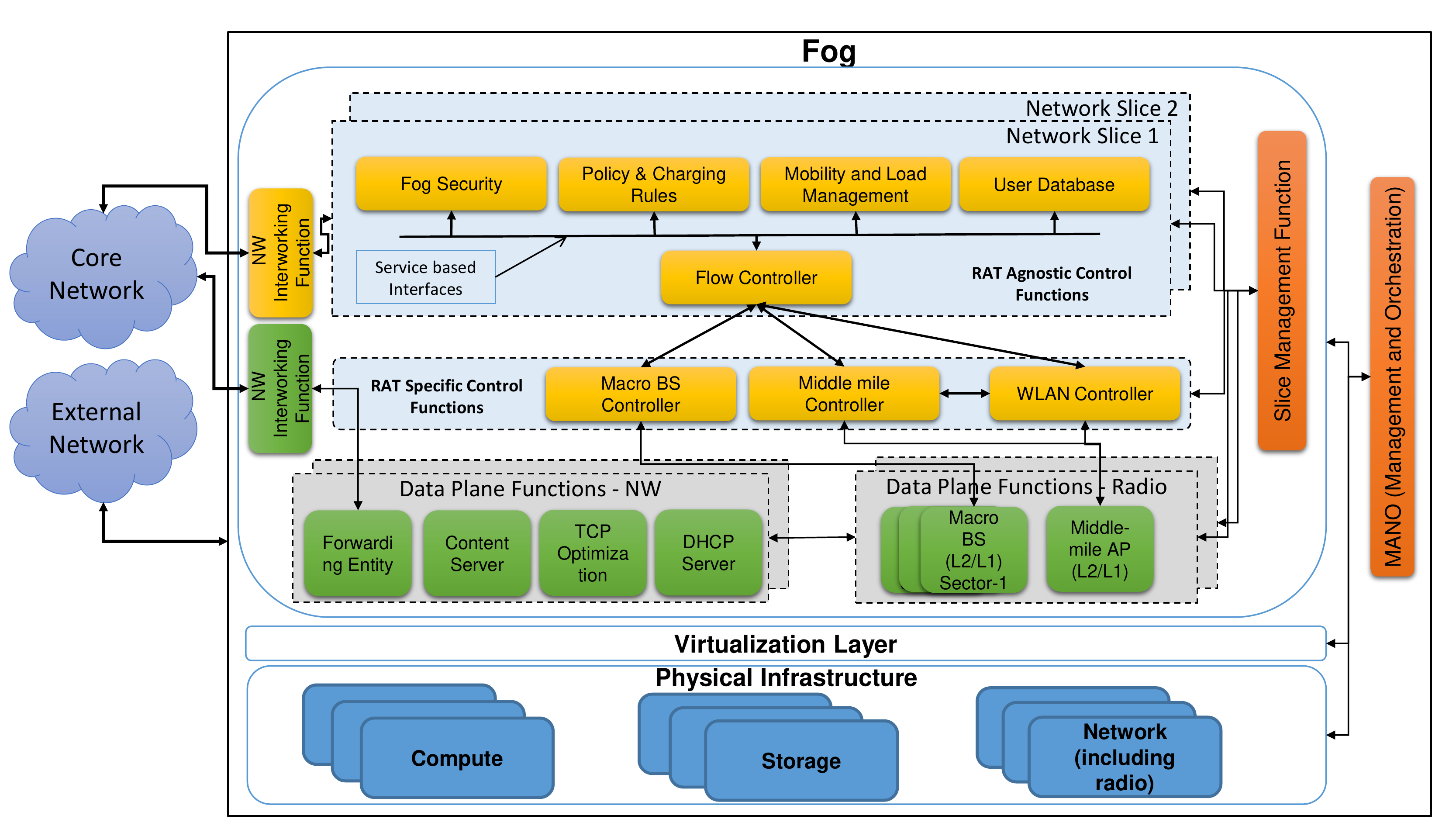}
		\caption{\label{fig:arch} Detailed architecture of Access Network (Fog)}
	\end{figure*}
	We now describe the primary building blocks of Frugal 5G Network Architecture as shown in Fig.~\ref{fig:arch}. We categorize all network functions in AN as well as CN under control plane or data plane functions. The CN is responsible for overall control of the Frugal 5G Network. The CN Control Plane performs the standard tasks, such as, mobility management, policy and charging control similar to the core network of other cellular systems. The CN sets up the traffic flow for a user through these data plane entities. They can also set up a path between two Fog elements, if required. Additionally, the CN control plane manages the distributed SDN controllers in each of the Fog elements. The data plane functions in the CN are responsible for forwarding the received data to another data plane entity in the network or to the external data network. The control and data plane functions across Fog and Cloud are synchronized with each other to avoid any inconsistencies. We now describe, in detail, the working of control and data plane in the AN. 
	
	\subsection{AN Control Plane}
	\label{sec:control}
	As shown in Fig.~\ref{fig:arch}, we propose a layered SDN controller in which the top level functions have a unified view of the Multi-RAT radio access network and are responsible for network-wide decision making. We refer to these functions as the \textit{RAT Agnostic Control Functions (RACFs)}. The decisions taken by the RAT agnostic functions are translated into RAT specific decisions by the lower level \textit{RAT Specific Control Functions (RSCFs)} and supplied to the corresponding data plane functions.
	\begin{itemize}
		
		    \item \textbf{RAT Specific Control Functions:} The RSCFs provide an abstract view of the underlying radio access network to the higher level control entities such as RACFs and the Slice Management Function (SMF). The abstract view of the resources enables a unified control of the Multi-RAT by the RACFs. It also enables virtualization of the radio access network and facilitates creation of multiple logical networks or network slices. The RSCFs perform the tasks such as device configuration, information broadcast, resource allocation and QoS management. It is important to note that these control functions primarily execute the decisions taken by the higher level RACFs. 
		    
			\item \textbf{RAT Agnostic Control Functions:}
			The Flow Controller is an RACF that directly interfaces with the RSCFs. It is responsible for allocating resources and establishment of the path for individual traffic flows through the network. It operates over the abstract resources provided by the lower level functions i.e. the RSCFs. The Flow Controller analyses the individual traffic flows and associates QoS attributes with each of them with the help of the Policy and Charging Rules Function (PCRF) and User Database Function (UDF). For instance, a traffic flow may be categorized as a real-time flow along with a stipulated guaranteed data rate. This flow related information may be provided to other RACFs such as the Mobility and Load Management Function (MLMF), which may utilize the information to perform the mobility and load management tasks. MLMF has a global view of all connected users which includes their location and the RAT association. UDF maintains the user profile for each of the subscribers, which is accessed to perform functions like user authentication and QoS determination. In traditional cellular networks, PCRF, UDF, and part of MLMF are located in the CN. However to enable localized communication under individual Fog elements, these functions can be  instantiated in the Fog as well, if the resources to house them in Fog are available. As we have moved sensitive user data from the Core to the edge of the network, we need to employ a Fog security function, which is responsible for secure access to the sensitive user data in Fog. 
			
			
			\item \textbf{Interfaces:} There are three type of interfaces in the proposed network. 1) The interactions between RACFs are enabled via service based interfaces and may use RESTful Application Program Interfaces (APIs)~\cite{onf} to get the required information from each other. 2) The interface between Flow Controller and the RSCFs is analogous to OpenFlow interface. OpenFlow protocol~\cite{onf} may need some modifications to be applicable in this scenario. 3) The interface between the RSCFs, e.g.,  Macro BS controller, or WLAN Controller  and the corresponding data plane entities may be analogous to F1-AP (a 3GPP defined interface between 5G gNB-Centralized Unit and gNB-Distributed Unit) or Control and Provisioning of Wireless Access Points (CAPWAP) protocol. 
			
			\item \textbf{Slice Management Function (SMF):} SMF creates multiple logical networks or network slices utilizing the abstract resource view provided by the RSCFs. The network slices may be formed across multiple operators or it may be service-based for a single operator. In rural areas, sharing of network infrastructure among multiple operators can considerably lower the network cost. The method of slicing the network resources is discussed, in detail, in Section~\ref{sec:mo}.     
		\end{itemize}

		\subsection{AN Data Plane:} 
		    The data plane in Fog broadly includes the data forwarding functions and service functions, such as, Dynamic Host Configuration Protocol (DHCP) Server. The functions instantiated in Fog include forwarding elements of Macro BS and middle mile AP. The forwarding elements of the WLAN APs and the middle mile clients are located in the village clusters. The forwarding elements of the three different RATs in the AN are controlled by their respective control functions instantiated in Fog. There are various service functions which are a part of the data plane in Fog such as DHCP Server, Content (cache) Server and Transmission Control Protocol (TCP) Optimization Functions. In traditional cellular networks, DHCP Server is located in the CN because CN is responsible for the session management and IP allocation. To enable local data flow without involving CN, a DHCP Server is instantiated in Fog. User experience may improve significantly by keeping these service functions near the users (in the distributed Fog elements) as they reduce the end-to-end latency. This also reduces the backhaul usage, thereby decreasing the operational cost of the network.

\section{Key Working Principles}
	In order to understand how Frugal 5G network can overcome connectivity challenges in rural areas, it is important to discuss the working principles of the Frugal 5G network. 
	\subsection{Wireless Backhaul Integration}
		The most important feature of Frugal 5G network is the integration of MMN in the access network. Typically, the cellular standards support relays which enable not more than two hops in the network. We envision the MMN to be a multi-hop mesh network. Therefore, integration of MMN within the Frugal 5G network is essential so as to deliver the required QoS to the end-users.  When a data packet has to be delivered to a user through the WLAN, the flow controller gives the required instructions to both the WLAN and the middle mile controllers and the required path is set up through the MMN and the WLAN. Note that, the MMN provides infrastructure for WLAN APs to communicate with the WLAN controller as well and the control information exchange between the WLAN controller and WLAN APs takes place over the same MMN, which carries the data for users.
		
	\subsection{Flexible Fog Control}
		The usage of NFV in AN (Fog) as well as CN (Cloud) allows for a flexible instantiation of functions across Fog and Cloud. The SDN Controller in the Cloud provides the policy configuration to the management and orchestration entity for instantiation of different network functions on the individual Fog elements. The AN architecture as shown in Fig~\ref{fig:arch} is one of the possible ways in which network functions can be instantiated. Their instantiation in the Fog may vary and is dependent on the availability of the resources. For instance, when the compute and storage resources are limited, then only a handful of network functions, e.g., radio specific data plane functions, forwarding functions, Core Network Interworking Function etc. are instantiated on a Fog element. On the other hand, if compute and storage resources are available but the backhaul bandwidth (network resources) is limited, a larger number of functions can be installed in the Fog, especially the functions which help reduce the backhaul usage, e.g., the Content Server.

	\subsection{Multi-operator Resource Sharing}
		\label{sec:mo}
	    Sharing network resources is an important technique that can considerably reduce the network cost. With the help of NFV and SDN, it is possible to split the physical network resources into multiple logical networks also called network slices. Each of these network slices or logical networks can be used for different purposes, e.g., the individual slices can be allocated to an operator through which the operator can serve the rural areas. It is important to understand how the network resources are split among multiple slices. Depending on the various parameters such as subscriber base, an operator might need a certain amount of network resources. As already mentioned, the RSCFs provide an abstract (virtual) view of the underlying network resources to the SMF. The SMF splits these virtual network resources into multiple network slices, each of which is allocated to individual operators. The RACFs are instantiated separately for each of these slices as they may be specific to the operator's network. This provides complete flexibility in terms of network resource sharing. When one operator's network (slice) experiences low load scenario, the resources can easily be diverted to other operator's network (slice), which may be experiencing higher load. 
		
		
    \subsection{Localized Communication Support}		
	
\begin{figure}
			\centering			
			\includegraphics[width=\linewidth,trim={0.4cm 1.4cm 2.9cm 1.7cm},clip]{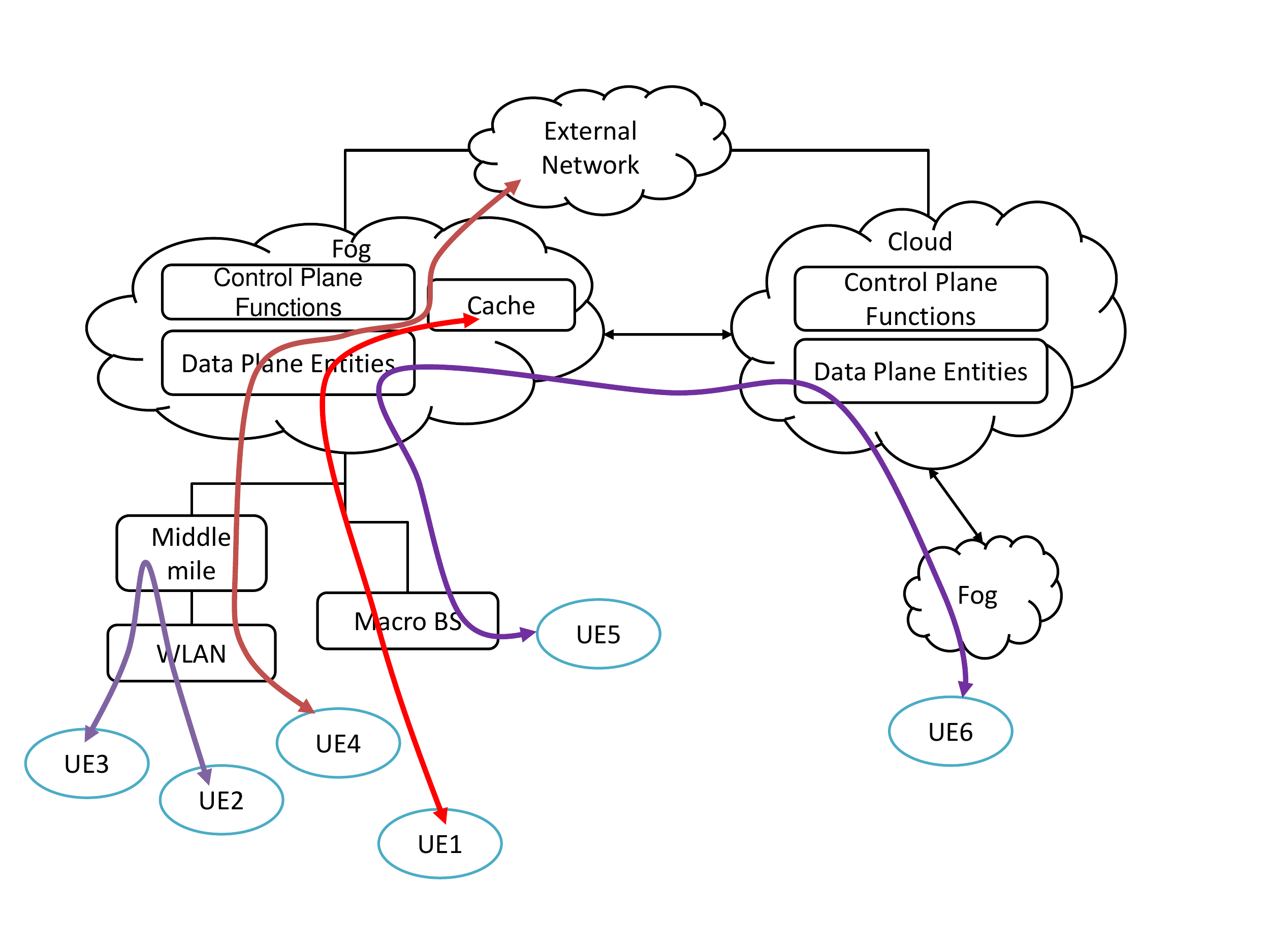}
			\caption{\label{fig:df} Data Flow Examples}
		\end{figure}

	In order to enable localized communication in the AN, without the involvement of Cloud, we have instantiated the relevant network functions in the Fog as explained in Section IV. This results in improvement of backhaul efficiency (by avoiding the Cloud-Fog signaling/data transfer) and user experience. To understand how the localized data paths are handled by Fog, it is important to understand the data/traffic flow set-up in the network. Various possible data/traffic flows in the network are illustrated in Fig~\ref{fig:df}. A group of SDN controllers in the ANs along with the CN is typically responsible for setting up the data path through the network for a user-specific traffic flow. Flow Controller in the AN sets up the data path through the data plane entities under its control. A data path for a user is selected either through a Macro BS or through the wireless MMN and a WLAN AP. Depending on various parameters, such as, the mobility of the user, the strength of the radio signal, QoS requirements, and load on different RATs, the flow controller selects an appropriate data path for a user. Now, it is possible that an end-to-end data path for a traffic flow is fully contained within a single Fog element. Example for such paths may be a Voice over IP (VoIP) session between two mobile users, who are in the close vicinity of each other (say, in the same village) or a data session between a user and the Content (cache) Server in the same Fog element. Usage of such data paths reduces the end-to-end latency of the flow and optimizes the resource utilization in the network by not using the wireless backhaul and the core Cloud elements. Since Fog elements can continue to provide services of localized nature in the absence of connectivity to the cloud, the network resilience is improved. Therefore, we may reduce the cost of network equipment by employing limited fault tolerance in the equipment.
		
\section{Open Issues and Challenges}
The work presented in this paper is an initial step towards designing the Frugal 5G network and it opens a lot of research issues and challenges which need to be addressed.
	\begin{itemize}
		\item \textit{Mobility management:} To successfully enable isolated operation of the Frugal 5G network, it is important to handle mobility management efficiently. There can be challenging scenarios where the SDN controller in Fog has set up the data path to facilitate communication among local users and one of the users has departed from the current Fog network. In such a scenario, how the session has to be maintained needs to be investigated. One method may be to allocate two IPv6 addresses to a user as done in distributed mobility management. 
		\item \textit{Network Interfaces:} The network interface protocol such as OpenFlow work only on the data flows. In the Frugal 5G network, there are flows in the network which are meant for user equipment such as those concerning authentication, admission control etc. OpenFlow will not be able to manage such flows. Some modifications are required in the OpenFlow protocol to facilitate such flows. 
		\item \textit{Middle mile network Planning:} We suggest a multi-hop MMN which serves a large rural area. Such a network requires optimized planning of the MMN so as to reduce the infrastructure cost of the network as well as improve its performance and coverage. Towards this, the development of a network planning tool is a principal necessity to perform efficient network planning in rural areas.
		\item \textit{Cache Management across Fog elements:} A Fog element has limited resources and therefore it can maintain a certain amount of cache in the network. It may not be sufficient to host all the relevant regional and popular content. One way to enable efficient content delivery is to host the popular content across various neighboring Fog elements. An important problem to study here is how the content is stored across the Fog elements and how many Fog elements can form a group to share the content.
	\end{itemize}
	
\section{Summary \& Conclusion}
    Current advances in cellular standards are driven by the requirements which are mostly urban in nature. Therefore, it is highly unlikely that next generation cellular standards will address the problem of broadband penetration in rural areas. To account for this, we have identified, in the paper, a set of traits which characterize the communication needs in rural areas. Based on the communication needs, we have proposed an innovative communication network architecture referred to as the Frugal 5G network. The proposed network architecture addresses the concerns through its unique and novel features---wireless backhaul RAN integration, the capability of isolated operation and localized communication, flexibility in service/function instantiation in the Fog and the framework for network virtualization and network slice creation. Table~\ref{tab:summary} summarizes how the Frugal 5G network architecture has addressed each requirement enumerated in Section \ref{sec:char}.
    
    \begin{table}[]
        \centering
        \caption{Summary of how Frugal 5G network architecture addresses the rural requirements}
        \renewcommand{\arraystretch}{1.2}
        \begin{tabular}{|p{3.5cm}|p{4.5cm}|}
            \hline
            \textbf{Requirements} & \textbf{Supporting Features in the Frugal 5G Network Architecture} \\ \hline \hline
            Affordability & low cost WLAN and middle mile nodes, wireless middle mile network, multi-operator sharing, flexible fog control \\ \hline
            Clustered service provision & high-speed WLAN connectivity only in clusters \\ \hline 
            Capable of running on renewable energy & low-power middle mile and WLAN nodes \\ \hline
            Facilitates local content generation and storage & cache enabled Fog element \\ \hline
            Limited mobility support & high-speed connectivity via WLAN backhauled by fixed wireless MMN \\ \hline
            Localized communication & local communication path set up via network functions instantiated in Fog \\ \hline 
            Device/equipment security & SDN enables centralized control and deployment of unintelligent nodes in the field reducing the security risk\\ \hline

        \end{tabular}
        
        \label{tab:summary}
    \end{table}
    
   We believe that our proposition in this paper can serve as a potential solution towards the IEEE P2061 standard. The development of the Frugal 5G network architecture opens up diverse research opportunities. For a start, some of the open issues and challenges have been identified in this paper which subsequent research will attempt to address.

	\bibliographystyle{ieeetr}
	\bibliography{myrefs}
	
\end{document}